\documentclass[preprint]{revtex4}
\usepackage{inputenc}
\usepackage{multirow,epsfig}
\usepackage{color}
\usepackage[english]{babel}
\usepackage{subfigure}
\usepackage{bbm}
\usepackage{graphicx}
\usepackage{amsmath}
\usepackage{amsfonts}
\usepackage{amssymb}
\usepackage{here}
\usepackage{setspace}
\def\gsim{\mathrel{\rlap{\lower4pt\hbox{\hskip1pt$\sim$}} \raise1pt\hbox{$>$}}} 
\def\lsim{\mathrel{\rlap{\lower4pt\hbox{\hskip1pt$\sim$}} \raise1pt\hbox{$<$}}} 

\begin{document}
\title{Calculation of electron-positron production in supercritical uranium-uranium collisions near the Coulomb barrier}
\begin{abstract}
The time-dependent Dirac equation was solved for zero-impact-parameter bare U-U collisions in the monopole approximation using a mapped Fourier grid matrix representation. A total of 2048 states including bound, as well as positive- and negative-energy states for an $N=1024$ spatial grid were propagated to generate occupation amplitudes as a function of internuclear separation. From these amplitudes spectra were calculated for total inclusive positron and electron production, and also the correlated spectra for ($e^+,e^-$) pair production. These were analyzed as a function of nuclear sticking time in order to establish signatures of spontaneous  pair creation, i.e., QED vacuum decay. Subcritical Fr-Fr and highly supercritical Db-Db collisions both at the Coulomb barrier were also studied and contrasted with the U-U results. 

\end{abstract}
\author{Edward Ackad and Marko Horbatsch}
\affiliation{Department of Physics and Astronomy, York University,\\ 4700 Keele St, Toronto, Ontario, Canada M3J 1P3}

\maketitle

\section{Introduction}
In heavy-ion collisions, dynamical electron-positron pairs can be created due to the time-varying potentials \cite{belkacemheavyion}. For a sufficiently strong static potential, pairs can also be created spontaneously. Such \textit{super-critical} potentials can be achieved in quasi-static adiabatic collision systems. The process of static pair creation is predicted by QED \cite{pra10324}, but has not yet been demonstrated unambiguously by experiment. Therefore, it continues to be of interest as a test of non-perturbative QED with strong fields. 

The Dirac equation is a good starting point to describe relativistic electron motion. As the nuclear charge, $Z$, of a hydrogen-like system increases, the ground state energy decreases. Thus, for a sufficiently high $Z$ value, the ground-state becomes embedded in the lower continuum, the so-called Dirac sea. The ground state changes character from a bound state to a resonant state with a finite lifetime, and is called a supercritical resonance state \cite{ackad:022503}. An initially \textit{vacant} supercritical state decays gradually into a pair consisting of a free positron and a bound electron. The bound electron will Pauli-block any other (spin-degenerate) pairs from being subsequently created. 

Currently, the only known realizable set-up capable of producing a supercritical ground state occurs in certain heavy-ion collision systems. When two bare nuclei collide near the Coulomb barrier, the vacant quasi-molecular ground state, i.e., the 1S$\sigma$ state, can become supercritical when the nuclei are sufficiently close. For the uranium-uranium system, the 1S$\sigma$ becomes supercritical at an internuclear separation of $R\lsim 36$ fm. As the nuclei continue their approach, the supercritical resonance experiences a shorter decay time. Thus, it is most probable that the supercritical resonance will decay at the closest approach of the nuclei.

Rutherford trajectories result in collisions were the nuclei decelerate as they approach and come to a stop at closest approach before accelerating away. Since the nuclei are moving very slowly at closest approach, pairs created at this time are due to the intense static potential rather than due to the nuclear dynamics. Nuclear theory groups have predicted that if the nuclei are within touching range the combined Coulomb potential may remain static (``stick") for up to $T=10$ zeptoseconds ($10^{-21}$s) \cite{EurPJA.14.191,zagrebaev:031602,Sticking2}. Such a phenomenon would enhance the static pair creation signal without changing the dynamical pair creation background.

In the present work, the time-dependent radial Dirac equation was solved in the monopole approximation for all initial states of a mapped Fourier grid matrix representation of the hamiltonian \cite{me1}. Single-particle amplitudes were obtained for an initially bare uranium-uranium head-on (zero-impact-parameter) collision at $E_{\rm CM}=740$ MeV, i.e, at the Coulomb barrier. The amplitudes were used to calculate the total positron and electron spectra for different nuclear sticking times, $T$. 

Previous work \cite{PhysRevA.37.1449,eackadconf2007} obtained the electron bound-state contribution to the total positron spectrum by following the time evolution of a finite number of bound-state vacancies. In the present work the complete correlated spectrum was calculated, by which we mean that the final ($e^+,e^-$) phase space contributions include both of (nS $e^-+$ free $e^+$) and (free $e^-+$ free $e^+$) pairs.

While the method is capable of handling any head-on collision system (symmetric or non-symmetric), the uranium system was chosen on the basis of planned experiments. Previous searches for supercritical resonances used partially ionized projectiles and solid targets \cite{PhysRevLett.51.2261,PhysRevLett.56.444,PhysLettB.245.153}. The ground state had a high percentage of being occupied, thus damping the supercritical resonance decay signal significantly due to Pauli-blocking. Over the next decade the GSI-FAIR collaboration is planning to perform bare uranium-uranium merged-beam collisions. A search will be conducted for supercritical resonance decay and for the nuclear sticking effect. Therefore, the present work will aid these investigations by providing more complete spectrum calculations. 

The limitation to zero-impact-parameter collisions is caused by the computational complexity. While this implies that direct comparison with experiment will not be possible, we note that small-impact-parameter collisions will yield similar results \cite{PhysRevA.37.1449}.

\section{Theory}
The information about the state of a collision system is contained in the single-particle Dirac amplitudes. They are obtained by expanding the time-evolved state into a basis with direct interpretation, namely the target-centered single-ion basis. These amplitudes, $a_{\nu,k}$, can be used to obtain the particle creation spectra \cite{PhysRevA.45.6296,pra10324}. 

The total electron production spectrum, $n_k$, and positron production spectrum, $\bar{n}_q$, where $k$ and $q$ label positive and negative  (discretized) energy levels respectively, are given by
\begin{eqnarray}\label{creat2}
\langle n_k \rangle &=& \sum_{\nu<F}{|a_{\nu,k}|^2} \\ \label{creat1}
\langle \bar{n}_q\rangle &=& \sum_{\nu>F}{|a_{\nu,q}|^2}.
\end{eqnarray}
Here the coefficients are labeled such that $\nu$ represents the initial state and $F$ is the Fermi level \cite{PhysRevA.37.1449,PhysRevA.45.6296}. Equations~\ref{creat2} and \ref{creat1} contain sums over all the propagated initial states above (positrons) or below (electrons) the Fermi level, which is placed below the ground state separating the negative-energy states from the bound and positive-energy continuum states. 

Therefore, all initial positive-energy states (both bound and continuum) must be propagated through the collision to calculate the total positron spectrum, which is obtained from vacancy production in the initially fully occupied Dirac sea. The dominant contribution (and a first approximation) to this spectrum is obtained by the propagation of the initial 1S state only \cite{eddiethesis}. M\"uller \textit{et al.} \cite{PhysRevA.37.1449} reported partial positron spectra for bare U-U collisions by propagating a number of initial bound-state vacancies.

Propagating all the (discretized) states is accomplished in the present work by solving the time-independent Dirac equation (for many intermediate separations) using a matrix representation. The wavefunction is first expanded into spinor spherical harmonics, $\chi_{\kappa,\mu}$, 
\begin{equation}
\Psi_{\mu}(r,\theta,\phi)=\sum_{\kappa}{ \left(
\begin{array}{c}
G_{\kappa}(r)\chi_{\kappa,\mu}(\theta,\phi) \\
iF_{\kappa}(r)\chi_{-\kappa,\mu}(\theta,\phi)
\end{array}
\right)} \quad ,
\end{equation}
which are labeled by the relativistic angular quantum number $\kappa$ and the magnetic quantum number $\mu$ \cite{greiner}. The Dirac equation for the scaled radial functions, $f(r)=rF(r)$ and $g(r)=rG(r)$, then becomes ($\hbar=$ c $=$ m$_{\rm e}=1$),
\begin{eqnarray}
\label{syseqnG}
\frac{df_{\kappa}}{dr} - \frac{\kappa }{r}f_{\kappa} & = & - \left(E -1 \right)g_{\kappa} + \sum_{\bar{\kappa}=\pm1}^{\pm\infty}{\langle \chi_{\kappa,\mu} \left| V(r,R) \right| \chi_{\bar{\kappa},\mu} \rangle }g_{\bar{\kappa}} \quad , \\
\label{syseqnG2}
\frac{dg_{\kappa}}{dr} + \frac{\kappa}{r}g_{\kappa} & = & \left(E + 1 \right) f_{\kappa} - \sum_{\bar{\kappa}=\pm1}^{\pm\infty}{\langle \chi_{-\kappa,\mu} \left| V(r,R) \right| \chi_{-\bar{\kappa},\mu} \rangle }f_{\bar{\kappa}} \quad ,
\end{eqnarray}
where $V(r,R)$ is the potential for two uniformly charged spheres displaced along the $z$-axis. This potential is expanded into Legendre polynomials according to $V(r,R)=\sum^{\infty}_{l=0}{V_l(r,R) P_l(\cos{\theta})}$ \cite{greiner} separating the equations into multipolar coupled equations. The mapped Fourier grid method is used to build a matrix representation of Eqs.~\ref{syseqnG} and \ref{syseqnG2}. Upon diagonalization of the matrix representation a complete basis is obtained spanning both the positive and negative continua \cite{me1}. 

Propagating the initial states is accomplished by the application of the propagator to each initial state. The full propagator is given by,
\begin{equation}\label{expH}
| \Psi(\mathbf{r},t)\rangle=\hat{{\rm T}}\left\{ \exp\left(-i \int_{t_0}^t dt' H(t')\right) \right\}| \Psi(\mathbf{r},t_0) \rangle,
\end{equation}
where $| \Psi(\mathbf{r},t_0) \rangle$ is the initial state, $\hat{{\rm T}}$ is the time-ordering symbol, and $H(t')$ is the time-dependent Dirac hamiltonian for the collision. Since the direct application of this propagator is not efficient for our purposes we use a short-time-step approximation: $H(t)$ is approximated to change linearly to $H(t+\Delta t)$. 

In order to propagate the initial state forward a time-step $\Delta t$, it is expanded into an eigenbasis of $H(t+\Delta t)$, with eigenvalues $E_m(t+\Delta t)$. The propagator applied to the $m$th state thus yields,
\begin{equation}\label{proptrap}
\hat{{\rm T}}\left\{ \exp\left(-i \int_{t}^{t+\Delta t} dt' H(t')\right) \right\} \approx  
\exp\left(-i \frac{E_m(t+\Delta t)-E_m(t)}{2}\Delta t \right).
\end{equation} 
This approximation to the integral is known as the trapezoid rule and has an error of $O(\Delta t^3)$. By solving for the basis at $t+\Delta t$, then projecting the state onto this basis and applying the approximate propagator in Eq.~\ref{proptrap}, a state can be propagated by $\Delta t$. Re-applying the method many times allows one to propagate the approximate eigenstates throughout the collision. It should be noted that this method allows one to propagate a large number of states (the complete discretized spectrum) efficiently.

\subsection{Correcting for the non-orthogonality of the basis}
The matrix representation of the hamiltonian generated by the mapped Fourier grid method is non-symmetric. Therefore, the eigenvectors do not have to be orthogonal. The mapping implies a non-uniform
coordinate representation mesh \cite{me1}, being dense at the origin but increasingly sparse as $r\rightarrow \infty$. Thus, for large $r$ the representation of Rydberg states and continuum states suffers from aliasing. We find that the lower-lying well-localized bound states satisfy orthogonality rather well (e.g., $\langle \phi_i|\phi_j\rangle \lsim 10^{-10}$ when $i\neq j$ for a calculation with $N=1024$), but that most continuum states are not really orthogonal. The deviation from zero in the orthogonality condition for continuum states can reach a value of 0.16 for $N=1024$ between continuum states near the ends of the continuum were the energies reach $10^6$ m$_e$c$^2$. For the relevant continuum states in this work, $E \lsim 20$m$_e$c$^2$, the deviation from zero is on the order of $10^{-4}$.

To correct for this deficiency the inverse of the inner-product matrix, $S_{i,j}=\langle \phi_i|\phi_j\rangle$, is needed. For a state 
\begin{equation}\label{psia}
\left| \Psi \rangle\right. = \sum_n{b_n \left| \phi_n \rangle\right.} 
\end{equation}
the inner-product coefficient is given by 
\begin{equation}\label{cm}
c_m = \langle \phi_m \left|\right. \Psi\rangle.
\end{equation}
Inserting \ref{psia} into \ref{cm} the expansion coefficient, $b_n$, is related to the inner-product coefficient according to
\begin{eqnarray}\label{expcoefdef}
c_m & = & \sum_n{b_n \left( \langle \phi_m \left|\right. \phi_n\rangle \right)} \\
\Rightarrow b_n & = & \sum_m{c_m \left( \langle \phi_m \left|\right. \phi_n\rangle \right)^{-1}}.
\end{eqnarray} 

The non-orthogonality of the mapped Fourier grid basis has no effect on the orthogonality properties of the Fock space basis. Thus, we can use the expressions in Eqs.~\ref{creat2} and \ref{creat1}. The new (orthogonal-basis) expansion coefficients, $b_n$, and not the inner-product coefficients are the single-particle amplitudes need for Eqs.~\ref{creat2} and \ref{creat1}. 

\subsection{Supercritical resonances}
Resonances are characterized by two parameters: the energy, $E_{\rm res}$, and the width, $\Gamma$, which is related to the lifetime, $\tau$ by $\Gamma = \hbar/\tau$. The interpretation for supercritical resonances is somewhat different from scattering resonances, e.g., in the case of a single supercritical nucleus ($Z\gsim 169$), a ground-state vacancy can decay into a ground-state electron and a free positron \cite{ackadcs}. The situation may be viewed as the scattering of the negative-energy electrons from the supercritical vacancy state. The behavior is then similar to electrons scattering off an atom in an external electric field, i.e., a standard atomic resonance. The decay of the resonance is characterized, in this case, by the ejection of a positron and the creation of a ground-state electron from the static QED vacuum. Thus, the interpretation of supercritical resonances differs from atomic resonances: the outgoing state represents a bound-free pair and not the release of a previously bound electron. Supercritical resonance decay is sometimes referred to as charged vacuum decay, since pairs of charged particles are created spontaneously \cite{rafelski}.

The basis obtained from the diagonalization of the hamiltonian generated by the mapped Fourier grid method contains the supercritical resonance \cite{ackadcs}. Thus, the effects of charged vacuum decay are included when the states are propagated through the collision with the current method: the positron spectrum includes the effects of supercritical resonance decay, as well as dynamical pair production.

Since nuclei with charges $Z\gsim 169$ do not exist one can resort to adiabatically changing Dirac levels (such as the 1S$\sigma$ quasi-molecular ground-state) in heavy-ion collisions near the Coulomb barrier \cite{pra10324}. As the nuclei approach, the quasi-molecular ground-state energy decreases continuously from its initial (subcritical) value to that of the supercritical resonance at closest approach. For a given combined charge $Z_1+Z_2$ a continuous range of supercritical resonance states will occur from the edge of the negative-energy continuum to the deepest-possible value at closest approach. As previously shown in Ref.~\cite{ackadcs,pra10324}, deeper-energy resonance states, have shorter life times. The resonance decay signal will be dominated by the deepest resonance, i.e. the resonance formed at closest approach, since the collision system spends most time at the Coulomb barrier, and also because $\tau$ decreases with $E_{\rm res}$ \cite{ackadcs,pra10324}. 

Previous calculations for spontaneous electron-positron production from bare heavy-ion collisions \cite{PhysRevA.37.1449} were performed within a computational basis with limited continuum energy resolution. While the present work is restricted to the zero-impact-parameter geometry, it is novel from the point of view of also propagating the discretized continua. This permits us to calculate not only all contributions to the total inclusive electron and positron spectra, but also the complete $(k,q)$ correlated spectra. Here $k$ labels any bound or unbound electron states, while $q$ represents the positron energy (or Dirac negative-energy) states.

Recently we developed methods for calculating resonance parameters including adding a complex absorbing potential (CAP) \cite{ackadcs}, complex scaling (CS) \cite{ackadcs} and smooth exterior scaling (SES) \cite{ackad:022503}. Using a CAP or SES, augmented with a Pad\'e approximant, highly accurate results for the resonance parameters were obtained with little numerical effort \cite{ackad:022503}. These have been used to parameterize the resonance ($E_{\rm res}$ and $\Gamma$) of the U-U collision, as well as a hypothetical Db-Db collision. The accurate knowledge of the decay times allows for a detailed comparison of these different supercritical systems, which is presented in section~\ref{t0sec}.

\subsection{Correlated spectra}\label{seccorr}
The positron and electron spectra obtained from Eqs.~\ref{creat2} and \ref{creat1} are the result of both dynamical and static pair creation processes. They contain no information about the state of the partner particle, i.e., they are inclusive. Dynamical positron production can lead to pairs in very different energy states. Static pair production can only produce pairs with a ground-state electron (assuming the field is not supercritical for excited bound states). The positrons emanating from the resonance decay are likely to have an energy close to the resonance energy. Positrons from dynamical (1S $e^-+$ free $e^+$) pairs will have a much broader spectrum determined by the collision's Fourier profile. Thus, the positron spectrum correlated to ground-state electrons should provide a cleaner signal of supercritical resonance decay than the inclusive $e^+$ spectrum.

The correlated spectrum is the sum of random (chance) correlations and true correlations. It is given by,
\begin{equation}\label{corr}
\langle n_k \bar{n}_q \rangle= \langle n_k\rangle \langle \bar{n}_q\rangle + \left| \sum_{j>F} a^*_{j,k}a_{j,q} \right|^2,
\end{equation}
where $k>F$ and $q<F$ \cite{PhysRevLett.43.1307}. Here the sum is taken over positive $e^+$ energy states, while in \cite{PhysRevLett.43.1307} it is expressed in terms of negative-energy initial states. The first term is interpreted as the random coincidence of a pair being in the $(k,q)$-state, while the second term is interpreted as the purely correlated spectrum, called $C_{k,q}$. Thus, the signal for the decay of the supercritical resonance will be primarily contained in the spectrum $C_{{\rm 1S}\sigma,q}$. For most situations the random contribution is small compared to the pure correlation term, since it involves the product of two small probabilities.


\section{Results}
The supercritical collision of two fully-ionized uranium nuclei was computed for a head-on Rutherford trajectory at a center-of-mass energy of $E_{\rm CM}=740$ MeV. All results are given in natural units ($\hbar=$ m$_{\rm e}=$ c $=1$): thus an energy of $E=\pm 1$ corresponds to $E=\pm$m$_{\rm e}$c$^2$, and an internuclear distance $R=1$ corresponds to $R=\hbar/($m$_{\rm e}$c$)\approx386.16$ fm. 

The calculations presented in this work were performed with a mesh size of $N=1024$ resulting in 2048 eigenstates \cite{me1}. The mapped Fourier grid scaling parameter value of $s=700$ was found to give stable results (cf. Ref~\cite{ackadcs}). The time mesh (in atomic time units where 1 a.t.u.$\approx 2.42\times10^{-17}$s) was broken up into three different parts with $\Delta t=4\times10^{-4}$ for $R>8$, $\Delta t=1\times10^{-4}$ for $8\geq R \geq 2$ and $\Delta t=1\times10^{-6}$ for $R < 2$.

For the U-U system a center-of-mass energy of 740 MeV corresponds to a closest-approach distance of 16.5 fm. This results in a deep resonance with $E_{\rm res}=1.56$ m$_{\rm e}$c$^2$ and $\Gamma=1.68$ keV calculated using the technique of Ref~\cite{ackad:022503}. The nuclei are assumed to be spherical with a radius of $R_{\rm n}=7.44$ fm (using $R_{\rm n}=1.2\times A^{1/3}$ with $A=$238 for uranium). The nuclei are thus 1.63 fm away from touching. This choice was made to facilitate comparison with previous calculations by M\"uller \textit{et al.} \cite{PhysRevA.37.1449}.

Additional collision systems were also considered to help understand the U-U results. The subcritical collision system of francium-francium ($A=222$) at $E_{\rm CM}=674.5$ MeV was chosen to analyze the purely dynamical processes. A dubnium-dubnium ($A=268$) collision at $E_{\rm CM}=928.4$ MeV was also chosen to better understand the supercritical resonance decay. The dubnium isotope, $A=268$ has a half-life of 1.2 days \cite{CRC}, perhaps making such experiments possible. Practical limitations, of course, exist as one would need to produce large numbers of ions in order to perform collision experiments. 

At closest approach the Db-Db system has resonance parameters $E_{\rm res}=3.57$ m$_{\rm e}$c$^2$ and $\Gamma=47.8$ keV, the latter giving for the lifetime of the resonance, $\tau=13.8$ zeptoseconds ($10^{-21}$ s). A strong signal should be detectable in this case since the decay time is only about an order of magnitude larger than the collision time. The collision energies were chosen to result in the nuclei being 1.63 fm from touching for each system. Thus the dynamics of all three systems is expected to be approximately the same. 

The data for the three systems are summarized in Table~\ref{tb1}. The resonance decay time $\tau$ should be compared to the dynamical time, $T_0$, which is the time when the 1S$\sigma$ state is supercritical (without sticking).
\begin{table}[htbp]
	\centering
\begin{tabular}{|c|c|c|c|c|c|c|}
\hline 
 Collision system & $Z_{\rm united}$ & $E_{CM}$ [MeV]& $E_{{\rm 1S}\sigma}$ [m$_{\rm e}$c$^2$]& $\Gamma$ [keV] & $\tau$ [$10^{-21}$s] & $T_0$ [$10^{-21}$s] \\
\hline\hline
Fr-Fr &  174 & 674 & -0.99 &	- & - & - \\
\hline
U-U &  184 & 740 & -1.56 &	1.68 & 392 & 2.3 \\
\hline
Db-Db &  210 & 928 &  -3.57 &	47.8 & 13.8 & 4.1  \\
\hline
\end{tabular}
	\caption{\label{tb1} Comparison of the parameters for the Fr-Fr, U-U and Db-Db systems at a closest-approach distance of near-touching (cf. text). For the supercritical U-U and Db-Db cases the positron resonance energy is obtained as $E_{res}=-E_{{\rm 1S}\sigma}$. For Fr-Fr the Dirac eigenvalue at closest approach is just above $E=-$m$_{\rm e}$c$^2$. The final column gives the time the 1S$\sigma$ state is supercritical along the Rutherford trajectory. }
\end{table}
It can be seen that for the Db-Db system the resonance decay time $\tau$ approaches the total supercritical time $T_0$.

\subsection{Collisions without sticking}\label{t0sec}
Propagating all initial (discretized) positive-energy states allows for the calculation of the total positron spectrum using Eq.~\ref{creat1}. The eigenstates of a single, target-centered nucleus were used for the initial conditions. The change of basis from a single nucleus to the quasi-molecular system in the target frame does not induce transitions provided the initial distance is far enough ($R \geq 25$).

\begin{figure}[H]
	\centering
		\includegraphics[angle=270,scale=0.3]{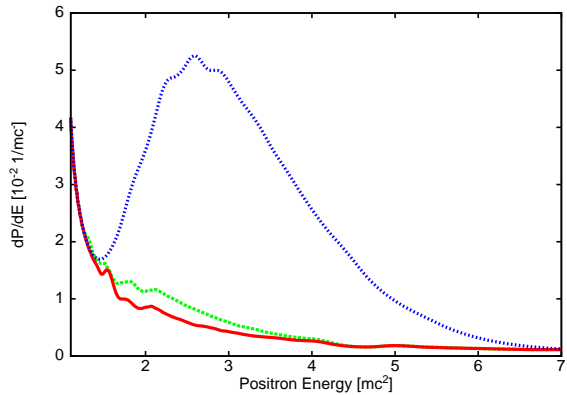}
	\caption{(Color Online) The total energy-differential positron probability spectrum for three collision systems at center-of-mass energies in which the nuclei are 1.63 fm from touching at closest approach. The dotted (blue) curve is for the Db-Db system which has a deep supercritical resonance (cf. Table~\ref{tb1}). The dashed (green) curve is for the U-U system under study. The solid (red) curve is for the Fr-Fr system which remains subcritical, providing a background spectrum resulting solely from the dynamics of the collision.}
	\label{fig1}
\end{figure}

Figure~\ref{fig1} shows the total energy-differential positron probabilities for the three collision systems.
The results for all three cases agree below $E<1.25$, which indicates that this part of the spectrum is purely dynamical. The Db-Db result shows a broad peak centered at $E \approx 2.6$. The U-U data show a deviation from the Fr-Fr results in the range $E=1.25-4.5$. The latter can be understood to represent the differential positron probability solely due to the dynamics of the system, without any effects from supercritical field decay. For positron energies where the Fr-Fr spectrum matches the other two, only positrons created by the changing potential in the collision are contributing. 

The data are suggestive of the fact that the spontaneous and dynamic contributions to the inclusive positron spectra are additive, although when comparing the Fr-Fr and Db-Db data one shouldn't rule out the possibility of an increase in dynamical positron production due to the dramatic change in $Z$, particularly at high positron energies.

The results for all systems are not entirely smooth. While the structures are stable with respect to change in calculation parameters such as the basis size, they do vary slightly with final separation. 

The Db-Db peak is not centered near $E_{\rm res}=3.57$, but at $E\approx 2.6$. In part this is due to the continuously varying intermediate supercritical resonance states before and after closest approach contributing in addition to the dominating closest-approach resonance \cite{PhysRevA.37.1449}. One can find how $\Gamma$ changes with $E_{\rm res}$ using data from different systems \cite{ackadcs}, and deduce that an energy peak of $E\approx 2.9$ is expected using
\begin{equation}\label{predictE} 
E_{\rm peak}\approx\frac{\int_{1}^{E_{\rm res}} \Gamma(E) E dE}{\int_{1}^{E_{\rm res}} \Gamma(E) dE}. 
\end{equation}
Thus, the interplay of the intermediate resonance state decays and the dynamical positrons (from the rapidly varying bound states) make a single sharp resonance peak at $E_{\rm res}$ unattainable without nuclear sticking. Note that a significant energy broadening is to be expected due to the collision time, $T_0$, being significantly shorter than the decay time of the resonance. Here $T_0$ is chosen as the time of supercriticality, and is given in Table~\ref{tb1}.

The dynamical background is almost exclusively due to the inclusion of positive-continuum initial states in Eq.~\ref{creat1}. Propagating only bound states yields a peak where the tails contain negligible differential probability, as was found in Ref.~\cite{PhysRevA.37.1449} based upon the propagation of bound vacancy states and their coupling to the negative-energy continuum. The low-lying positive-continuum states in Fig.~\ref{fig1} show significant positron production probabilities. These states are excited due to their necessity in representing the quickly changing (super- or sub-critical) ground state in the asymptotic basis.

\begin{figure}[H]
	\centering
		\includegraphics[angle=270,scale=0.3]{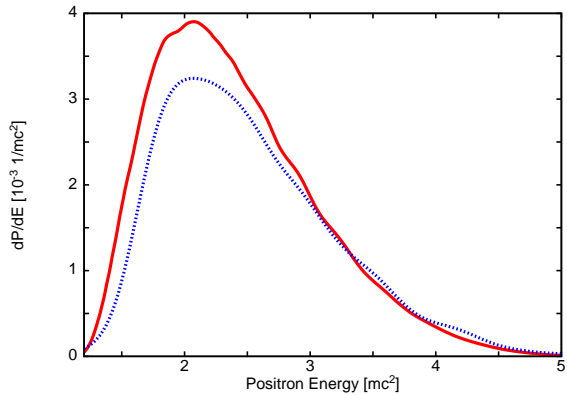}
		\caption{(Color Online) The difference between the U-U and Fr-Fr total differential positron spectra shown in Fig.~\ref{fig1} is represented by the dashed (blue) curve. The solid (red) curve shows the correlated differential spectrum, $C_{{\rm 1S},q}$. }
	\label{fig2}
\end{figure}

The subcritical dynamical background may be subtracted from the U-U spectrum, highlighting the effects of the supercritical resonance decay on $dP/dE$. This is done by subtracting the Fr-Fr spectrum from the U-U spectrum. The result is the dashed (blue) curve shown in Fig.~\ref{fig2}.
The decay signal is heavily smeared by the dynamical process leading into, and out of, supercriticality. Thus, a sharp resonance peak with a width $\Gamma$ is not observed. The spectrum still includes dynamical effects of the system being supercritical with the ground state inside the negative-energy continuum. Nonetheless, the resultant spectrum shows a single peak centered at $E\approx2.1$, close to $E=1.8$ predicted by Eq.~\ref{predictE}.

This difference spectrum can be compared with the correlated (1S $e^-$, free $e^+$) spectrum, $C_{{\rm 1S},q}$, which is shown as the (red) solid curve in Fig.~\ref{fig2}. This correlated spectrum was calculated using the second term of Eq.~\ref{corr} and provides a more distinct signal for supercritical resonance decay, as can be seen by comparing Fig.~\ref{fig1} and \ref{fig2}. All the states of the finite representation spanning the two continua and the bound states were used. 

The correlated spectrum is also peaked at $E\approx 2.1$ and matches the inclusive spectrum at low and high energies. The difference peak is slightly below the correlated peak. This is unexpected as the correlated spectrum contains fewer processes, such as  other (nS $e^-$, free $e^+$) pair production when the system is supercritical, predominantly dynamical (2S $e^-$, free $e^+$) production. If there was no change in the dynamical contributions to the inclusive positron spectra when going from $Z_{\rm united}=174$ to $Z_{\rm united}=184$, and if the additivity of the dynamical and supercritical positron productions was perfect, then one would expect the inclusive difference spectrum to exceed somewhat the $C_{{\rm 1S},q}$ correlated U-U spectrum. Thus, the comparison shows that a proper isolation of the supercritical contribution in the inclusive spectra requires a determination of how the dynamical contributions scale with $Z$. 

The coincidence ($n$S $e^-$, free $e^+$) spectra can be compared to previous calculations where only a few ($n$S) vacancies were propagated to obtain a partial inclusive spectrum \cite{PhysRevA.37.1449}. We find very good agreement in the case of (1S $e^-$, free $e^+$) pair production, and an overestimation of the correlated spectra by the simpler calculation in the case of the (2S $e^-$, free $e^+$) channel.

\subsection{Effects of nuclear sticking on uranium-uranium collisions}
The phenomenon of nuclear sticking enhances the supercritical resonance decay signal. Trajectories with nuclear sticking are obtained by propagating the states at closest approach while keeping the basis stationary for a time, $T$. No basis projections are needed for this part, since the basis remains static and the time evolution of each eigenstate is obtainable from a phase factor. 

\begin{figure}[H]
	\centering
		\includegraphics[angle=270,scale=0.3]{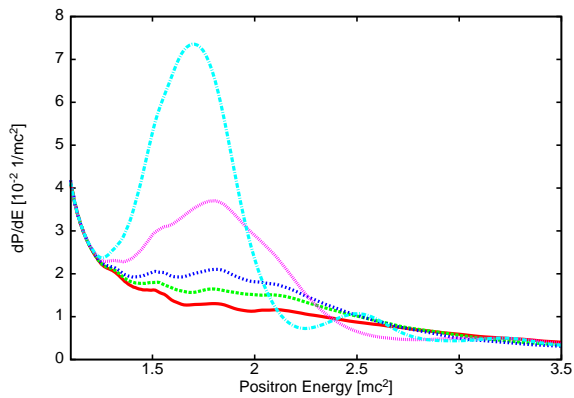}
	\caption{(Color Online) The total differential positron spectrum, $dP/dE$ for the U-U system at $E_{CM}=740$ MeV, for different nuclear sticking times calculated using Eq.~\ref{creat1}. The solid (red) curve is for $T=0$, the long dash (green) for $T=1$, the short dashed (blue) for $T=2$, the dotted (magenta) for $T=5$ and the dashed-dot (cyan) for $T=10$. }
	\label{fig3}
\end{figure}

Figure~\ref{fig3} shows $dP/dE$ for $T=0,1,2,5$ and 10 zeptoseconds calculated from Eq.~\ref{creat1}. A sticking time of 10 zeptoseconds is the longest time predicted as realistically attainable by the nuclear theory groups \cite{EurPJA.14.191,zagrebaev:031602,Sticking2}. All initial positive-energy states were propagated in order to compute the inclusive spectrum. 
As $T$ increases the peaks become narrower and move closer to the closest-approach resonance position of $E_{\rm res}=1.56$. All curves converge for $E<1.25$ and are close for $E>4.2$. The secondary peaks exhibited by the $T\geq 5$ curves, in the range displayed, are at $E=2.5$ for $T=10$ and $E=3.25$ for $T=5$.

The decay of the supercritical resonance also results in an increase in the total number of positrons created in conjunction with electrons in the ground state. Most of these electrons remain in the ground state, while in a few cases they are subsequently excited into higher states. Table~\ref{table} gives the probabilities for the three lowest bound states calculated using Eq.~\ref{creat2} including all positron continuum states.
\begin{table}[htbp]
	\centering
\begin{tabular}{|c|c|c|c|}
\hline 
 $T$ & 1S & 2S & 3S \\
\hline\hline
0 & 0.633 & 0.0225 &	7.38$\times10^{-3}$\\
\hline
1 & 0.916 & 0.0354 & 1.10$\times10^{-2}$ \\
\hline
2 & 1.16 & 0.0475 & 1.30$\times10^{-2}$ \\
\hline
5 & 1.95 &  0.0321 & 9.39$\times10^{-3}$ \\
\hline
10 & 3.08 & 0.0958 & 3.58$\times10^{-2}$ \\
\hline
\end{tabular}
	\caption{\label{table} Final electron creation probabilities of the low-lying bound states for U-U at $E_{CM}=740$ MeV for different nuclear sticking times, $T$. These were calculated by evolving all positron continuum states of a matrix representation of the hamiltonian using Eq.~\ref{creat2}. }
\end{table}
The increase in the probability of the bound states shown from $T=0$ to just $T=1$ is over 45\%. The probabilities increase further as $T$ increases in all cases except for the excited states in the case of $T=5$. For this collision we observe a decrease in the population of the excited states. It does not happen for all of the exicted states. We suspect that for a sticking time of about $T=5$ the dynamical interference which causes the many secondary peaks in the positron spectrum is responsible for this behavior. Thus, the 2S and 3S populations are lower due to destructive interference with dynamical processes. 

There is a particularly strong increase in the ground-state population as $T$ increases. An almost five-fold increase is observed when increasing the sticking time from $T=0$ to $T=10$. This is primarily due to the supercritical resonance decay. The average charge state of the uranium ions after the collision can thus be used as an aid in the detection the supercritical resonance decay by using a coincidence counting technique.

The continuum electron spectrum was also calculated using Eq.~\ref{creat2}. The results are the same for all $T$, and follow the same curve as the $T=0$ positron spectrum in Fig.~\ref{fig3}. Thus, the free electrons are not influenced by the sticking time and the statically created ground-state electrons are very unlikely to be excited to the continuum during the outgoing part of the collision.

\begin{figure}[H]
	\centering
	\includegraphics[angle=270,scale=0.3]{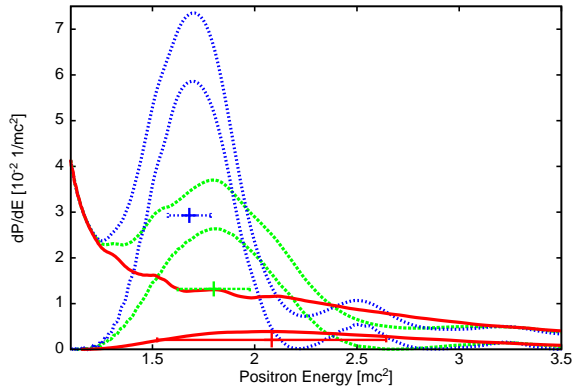}
	\caption{(Color Online) The total positron spectrum for the U-U system at $E_{CM}=740$ MeV for different nuclear sticking times (upper curves) compared to the correlated spectrum of positrons with partner electrons in the 1S state (lower curves with the same color). The red curves are for $T=0$, the (green) dashed curves are for $T=5$ and the (blue) dotted curves are for $T=10$. The correlated positron spectra contain only a small amount of dynamical positron background. The points with horizontal error bars of the same color represent the width estimate ($\hbar/(T+T_0)$ with $T_0=2.3\times10^{-21}$s, cf. Table~\ref{tb1}). }
	\label{corpos}
\end{figure}

The increase in the 1S population, as well as the increase in the main peak in the positron spectrum are all indirect signs of supercritical resonance decay. As shown above, a more direct signal is given by the correlation spectrum. Figure~\ref{corpos} shows the total differential positron spectra along with the differential correlation spectrum, $C_{k,q}$, using Eq.~\ref{corr} with $k=$1S. The $T=0$ system is shown as a solid (red) curve, the $T=5$ as a dashed (green) curve and the $T=10$ as a dotted (blue) curve. The correlated spectra are in the same style (color) as the total differential spectra of the same sticking time (the lower curve representing the correlated spectrum). This is due to the spectrum being selective about which positrons are included, i.e., only positrons created together with an electron in the ground-state (either the 1S or 1S$\sigma$ for large $R$). 
The correlated peaks also become narrower and are centered closer to $E_{\rm res}$ as $T$ increases in the same manner as $dP/dE$.  The correlated spectra resemble closely the results from propagating the bound states only, as done in Ref.~\cite{PhysRevA.37.1449}. They lack the significant dynamical background, which can be estimated from the Fr-Fr collision system (cf. Fig.~\ref{fig1}). Thus, the correlated spectra are dominated by the supercritical resonance decay, i.e. the desired signal. In experiments one can aim, therefore, for a comparison of inclusive positron spectra with those obtained from a coincidence detection where one of the ions has changed its charge state.

For $T\geq5$ additional peaks are observed in the correlated spectra. The $T=10$ spectrum exhibits a secondary peak at $E \approx 2.6$, and a third smaller peak, in the range shown, at $E\approx3.25$. The $T=5$ result has a single secondary peak in the range shown, coincidentally also at $E\approx 3.25$. The second peak for $T=5$ is much broader than both the second and third peaks in the $T=10$ correlation spectrum. 

The secondary peaks are found for the trajectories including sticking. The sticking changes the dynamics of the collision, and therefore causes the peaks. These peaks were also observed in the Fr-Fr system when sticking was included, eliminating the possibility that they are due to the supercriticality of the system. The energy separation between the peaks was found to depend only on $T$. The secondary peaks have also been previously seen in less energetic U-U collisions where $E_{CM}=610$ MeV \cite{eackadconf2007}. The series of peaks begin with the second peak and decrease in amplitude with peaks extending to energies above 50 m$_{\rm e}$c$^2$. In all systems studied, it was found that this series of secondary peaks had separations that were largely independent of the system. Thus, the decay of the supercritical resonance adds to the amplitude of the secondary peaks, but does not cause them.

At half of the respective peak's height in Fig.~\ref{corpos}, a width estimate of the resonance decay spectrum $\hbar/(T+T_0)$ is shown in the same color as the curve. Here $T_0=2.3$ zeptoseconds, i.e., the time the system is supercritical without sticking, was chosen to estimate the supercritical decay without sticking. The comparison shows that the width of the dominant peak structures are not explained perfectly. The estimate ignores the interplay between dynamical and spontaneous positron production. Better agreement would be obtained for smaller $T_0$ which can be justified by the observation that supercritical resonance decay becomes appreciable only when the 1S$\sigma$ state is embedded more deeply in the $E<-$m$_{\rm e}$c$^2$ continuum.

The correlation spectrum, $C_{k,q}$, defined by Eq.~\ref{corr} can also be calculated for (free $e^-$, free $e^+$) correlated pairs. This type of pair production can be detected more easily. For all systems described in this paper, none had a discernible signal. This is not surprising due to the intermediate collision energies, and the fact that static pair creation does not lead to free electrons. 

Using Eqs.~\ref{creat2}, \ref{creat1} and \ref{corr} it is possible to examine the spectra immediately following supercriticality. The resultant spectra in the target-frame two-center basis, would be indistinguishable from those shown in Figs.~\ref{fig3} and \ref{corpos}. There is almost no change in the total positron spectrum. Thus, almost all the positrons are created once the system emerges from supercriticality. Similarly, the correlated spectrum does not change. After supercriticality no static pairs can be created, and we find that dynamic pair creation ceases to be effective. 

\section{Conclusions}
Total inclusive positron and electron spectra were calculated for U-U zero-impact-parameter collisions at $E_{CM}=740$ MeV with nuclear sticking times $T=0,1,2,5$ and 10 zeptoseconds. The present calculations, while limited in geometry, represent a first attempt at propagating a dense representation of both positive- and negative-energy  eigenstates. Correlated spectra were also calculated and results were presented for (1S $e^-+$ free $e^+$) pair creation, the only channel available for supercritical decay in this system. While we found that the (1S $e^-+$ free $e^+$) correlated spectra agreed rather well with naively calculated spectra from propagating a 1S vacancy, it is important to note that substantial differences were observed for the case of (2S $e^-+$ free $e^+$) pairs.

The results were compared to subcritical Fr-Fr and highly supercritical Db-Db collisions at the Coulomb barrier. It was found that the correlation spectra provide the clearest signal for the supercritical resonance decay, although one can subtract the inclusive positron spectra to isolate the supercritical positron production from the purely dynamical contributions. This was demonstrated by the fact that at low and at high energies the inclusive spectra merge for the three systems: U-U, Fr-Fr and Db-Db. This differs from previous findings of Ref.~\cite{pra10324}, where a constant $E_{CM}$ was used, whereas the present work employed a constant internuclear touching distance by changing $E_{CM}$.  

The secondary peaks found in previous work \cite{eackadconf2007,PhysRevA.37.1449} were explained as resulting from the change in the trajectory due to nuclear sticking. They were found in subcritical collisions, eliminating the possibility that they were related to supercritical resonance decay. 

Some bound-state population results were also presented. The results show large increases in the bound-state populations as the nuclear sticking time $T$ increases. Given that near-zero-impact-parameter collisions can be selected by a nuclear coincidence count (large scattering angles), this channel will provide a strong indication of the nuclear sticking effect. Thus, an increase in the average charge state of the nuclei, together with sharper positron peaks, could be used to demonstrate the existence of long sticking times and the existence of charged vacuum decay.

The calculations presented in this paper reported single-electron Dirac calculations with spin degeneracy. Thus, for small pair production probabilities it is correct to multiply them by a factor of two. For the (1S $e^-$ , free $e^+$) correlated spectra one would have to take into account Pauli blocking effects and make use of many-electron atomic structure once the probabilities become appreciable.

The current work was performed in the monopole approximation to the two-center Coulomb potential. In previous work where static resonance parameters were calculated it was shown that the next-order effect is caused by the quadrupole coupling of the S$-$D states \cite{eackadconf2007}. While this coupling will certainly modify the resonance, e.g., increasing its energy, $E_{\rm res}$, it is unlikely that it will change the spectra significantly. This conclusion is based on the observation that substantial changes in $\Gamma$ due to higher-order contributions occur at larger internuclear separations where $\Gamma$ is small. It is expected that the main peak will be somewhat higher and broader, as indicated by the Db-Db results, due to the increased width $\Gamma$. 

\section{Acknowledgments}
The authors would like to thank Igor Khavkine for useful discussions. This work was supported by NSERC Canada, and was carried out using the Shared Hierarchical Academic Research Computing Network (SHARCNET:www.sharcnet.ca). E. Ackad was supported by the Ontario Graduate Scholarship program.

\bibliography{collisionUU}
\end{document}